# Title: Ultranarrow-linewidth Wavelength-Vortex Metasurface Holography

**Authors:** Weijia Meng[1,2]†, Johannes E. Fröch[3,4]†, Ke Cheng[1,2], Baoli Li[1,2], Arka Majumdar[3,4], Stefan A. Maier[5,6], Haoran Ren[5*], Min Gu[1,2*], and Xinyuan Fang[1,2*]

**Affiliations:**

[1]School of Artificial Intelligence Science and Technology, University of Shanghai for Science and Technology, Shanghai, 200093, China.

[2]Institute of Photonic Chips, University of Shanghai for Science and Technology, Shanghai, 200093, China.

[3]Department of Physics, University of Washington, Seattle, WA 98195, USA.

[4]Department of Electrical and Computer Engineering, University of Washington, Seattle, WA 98195, USA.

[5]School of Physics and Astronomy, Monash University, Melbourne, VIC 3800, Australia.

[6]Department of Physics, Imperial College London, London SW7 2AZ, UK.

*Corresponding author. Email: xinyuan.fang@usst.edu.cn (X.F.); haoran.ren@monash.edu (H.R.); gumin@usst.edu.cn (M.G.)

†These authors contributed equally to this work.

**Abstract:** Ultrathin metasurface holograms, with thicknesses comparable to the operating wavelength, leverage multiple degrees of freedom of light to address independent image channels, thereby significantly enhancing information capacity. Although the wavelength of light can be used to encode holographic image channels, high-capacity wavelength-multiplexing holography has traditionally been achieved only through 3D volume holograms based on Bragg diffraction. We demonstrate ultranarrow-linewidth wavelength-vortex multiplexing holography in ultrathin metasurface holograms. By applying dispersion engineering to the elementary grating functions of a multiplexing hologram, we develop a sparse $k$-vector-filtering aperture array in momentum space that achieves sharp wavelength selectivity in conjunction with orbital angular momentum selectivity. Further leveraging transformer neural networks for the design of phase-only multiplexing holograms, we reconstruct up to 118 independent image channels from a single metasurface hologram, achieving an ultranarrow linewidth of 2 nm in the visible range. Finally, we apply the developed wavelength-vortex multiplexing metasurface holograms for holographic visual cryptography, achieving unprecedented security with an information rate more than 2500 times higher than that of traditional visual cryptography schemes. Our results open exciting avenues for the use of metasurface holograms in various applications, including 3D displays, holographic encryption, beam shaping, LiDAR, microscopy, data storage, and optical artificial intelligence.

**Main Text:** Holography is a powerful technology for reconstructing the amplitude and phase information of 3D objects(*1*), with applications in both fundamental research(*2-6*) and the entertainment industry(*7*). To enhance hologram information capacity(*8*), various physical degrees of freedom of light have been utilized to carry independent image channels for multiplexing(*9-18*). The wavelength of light, plays a crucial role in optical multiplexing for displays(*19*), optical communications(*20*), hyperspectral imaging(*21*), and computing(*22, 23*), as well as in human visual perception(*24*). However, high-capacity wavelength multiplexing holography can currently only be recorded by a 3D volume hologram with strict *k*-vector selectivity based on Bragg diffraction(*25*).

In a volume hologram, as shown in Fig. 1A, independent elementary holographic output *k*-vectors ($k_{out1}$ and $k_{out2}$) encoded at different wavelengths are stored as thick gratings ($k_{g1}$ and $k_{g2}$) with a thickness much larger than the light wavelength. Owing to Bragg diffraction, only light with specific wavelengths and angles can be diffracted efficiently, resulting in sharp *k*-vector (wavelength) selectivity in volume holograms. Nevertheless, designing and fabricating thick volume holograms with specific materials remains a significant challenge(*9*), further complicated by the lack of multi-dimensional light manipulation (e.g., polarization(*17*) and vortex(*14*) sensitivities) and the compactness required for modern applications such as VR/AR(*26*), holographic displays(*27*), and optical computing(*28*).

Ultrathin metasurfaces, composed of a planar array of subwavelength structures with thicknesses comparable to the operating wavelength, offer a versatile platform in information optics, capable of shaping the wavefront of light across multiple degrees of freedom, including amplitude(*29*), phase(*30*), polarization(*31*), wavelength(*10*), and orbital angular momentum (OAM)(*32*). Metasurface holograms, or meta-holograms, provide exceptional advantages, including flexible operation bandwidth, a large space-bandwidth product, and integration with other optoelectronic devices without undesired diffraction orders(*33*). High-capacity meta-holograms have been demonstrated using polarization-(*17, 34-36*) and OAM-multiplexing(*37-39*) techniques. However, wavelength-multiplexing meta-holograms, achieved through multi-resonance meta-atoms(*40*), interleaving metasurfaces(*41-44*), and wavelength-dependent grating shifts(*11, 45*) suffer from low multiplexing bandwidth due to the lack of sharp wavelength selectivity.

Here we demonstrate ultranarrow-linewidth wavelength-vortex metasurface holography through elementary dispersion engineering of *k*-vector components in multiplexed image channels. Unlike volume holograms, we show for the first time that high-capacity wavelength multiplexing can be achieved using ultrathin metasurfaces. To achieve sharp wavelength selectivity, we engineer the dispersion of each *k*-vector component in wavelength-multiplexed image channels (Fig. 1B). Specifically, wavelength-dependent grating functions, denoted as $k_{g1}$, $k_{g2}$, with the periods satisfying $\lambda_1/\Lambda_1=\lambda_2/\Lambda_2$, guide light of desired wavelengths into the same diffraction angle $\beta_0$. We then use a sparse *k*-vector-filtering aperture array (KAA) in the momentum space of holograms to select desired holographic output *k*-vectors ($k_{out1}$ and $k_{out2}$) encoded by the correct wavelengths ($\lambda_1$ and $\lambda_2$). Notably, this KAA effectively blocks undesired wavelengths ($k_{out3}$) and spurious diffraction orders ($k'_{out1}$ and $k'_{out2}$), resulting in an ultranarrow linewidth of 2 nm for high-capacity wavelength-multiplexing metasurface holography in the visible range (Fig. 1C). Since the superposition of grating functions leads to complex-amplitude holograms, we adopt a transformer neural network to convert the complex-amplitude modulation, typically challenging for experimental implementation, into phase-only holograms. Finally, we demonstrate holographic visual cryptography (HVC) based on the combined use of wavelength and OAM degrees of freedom for shared secret codes (Fig. 1D). Due to the ultranarrow linewidth of the incident beams, our approach achieves unprecedented security in HVC, where concealed colorful image information can only be revealed from a metasurface ciphertext (meta-ciphertext) based on the holographic reconstruction of all the necessary monochromatic secrets using correct wavelength-vortex beams.

**Design principle**

The physical process to obtain such a high-capacity wavelength-vortex multiplexing hologram is illustrated in Fig. 2. A light beam with a specific wavelength-vortex state is utilized as a virtual reference beam to interfere with the object beam at the recording plane in the spatial frequency domain (Fig. 2A). To apply the KAA mentioned above and thus obtain sharp wavelength selectivity, a 2D Dirac comb function with the same sampling constant is adopted to discretize all target images. As a result, the intensity distribution of the interference pattern ($I(x, y)$) can be expressed as

$$I(x,y) = |C_1|^2 + |C_2|^2 + A^* \exp(jl\varphi) \sum_{m=-\infty}^{+\infty} \sum_{n=-\infty}^{+\infty} c_{m,n} \exp\left[-j2\pi(\frac{mD}{\lambda f}x + \frac{nD}{\lambda f}y)\right] \\ + A\exp(-jl\varphi) \sum_{m=-\infty}^{+\infty} \sum_{n=-\infty}^{+\infty} c_{m,n} \exp\left[j2\pi(\frac{mD}{\lambda f}x + \frac{nD}{\lambda f}y)\right],$$

(1)

wherein the first two terms of the above equation represent the intensity distributions of the object beam and the reference beam, respectively. The third and fourth terms, in conjugation to each other, represent the wavelength-vortex hologram. We use the third term in Eq. 1 as our digital hologram, where $c_{m,n}$ denotes the Fourier coefficient, $\lambda$ represents the encoded wavelength, $f$ is the focal length of the Fourier lens, $D$ represents the sampling constant, and ($mD$, $nD$) denote the orthogonal coordinates of the sampling array, respectively. To preserve the OAM information in each pixel of the target image, the sampling constant is determined by the maximum spatial frequency ($d$) of the reference beam (Fig. 2A). It is worthwhile noting that the periods of constituent grating functions are given as $\Lambda_x=\lambda f/mD$ and $\Lambda_y=\lambda f/nD$, satisfying the requirement of guiding the desired wavelengths into the same diffraction angles with $f/D$ as a constant. The superposition of the wavelength-vortex holograms comprising wavelength-dependent grating functions and encoding vortex phase plates results in a wavelength-vortex multiplexing hologram (Fig. 2B), which can be mathematically expressed as

$$E(x,y) = \sum_{p=1}^{P} \sum_{q=1}^{Q} \sum_{m=-\infty}^{+\infty} \sum_{n=-\infty}^{+\infty} c_{p,q,m,n} \exp\left[-j2\pi(\frac{mD}{\lambda_p f}x + \frac{nD}{\lambda_p f}y)\right] \exp(jl_q\varphi).$$

(2)

Therein, $P$ and $Q$ represent the total number of encoded wavelength and OAM channels. Notably, phase-only holograms are usually adopted due to its high efficiency and ease of fabrication. However, the loss of amplitude information in the wavelength-vortex multiplexing hologram can result in image crosstalk (*39, 46, 47*). To mitigate this issue, we develop a transformer neural network to obtain the Fourier coefficients of a phase-only wavelength-vortex multiplexing hologram(*48*) (fig. S1).

Through the above design principle, we demonstrate ultranarrow-linewidth wavelength-vortex multiplexing holography. An ultrahigh-capacity wavelength-vortex multiplexing hologram is designed to display the whole periodic table, based on twelve OAM modes with indices ranging from -12 to 12 at intervals of two and ten different wavelengths in the range between 413 nm and 700 nm. While spurious and undesired diffraction orders could result in crosstalk between different wavelength information channels with an average SSIM of ~0 (fig. S2), 118 independent images with high fidelity can be successfully reconstructed after imaging filtering by the KAA (Fig. 2C), wherein the SSIM is boosted to ~1 (upper layer of Fig. 2D). Without losing generality, the linewidth throughout this work is defined as the interval of two adjacent wavelengths when average SSIM drops to 0.2. By continuously changing the incident wavelength, we analyzed the SSIM (lower layer of Fig. 2D), exhibiting the linewidth variation from 2 nm to 11 nm (Fig. 2E). This suggests that deviation from the desired wavelengths and the OAM states of incident light will result in the quality degradation of reconstructed images (fig. S3).

**Ultranarrow-linewidth Wavelength-vortex Multiplexing Meta-holography**

A broadband geometric metasurface is used to implement our designed ultranarrow-linewidth wavelength-vortex multiplexing hologram (*30*). The meta-hologram is constructed from an array of subwavelength silicon nitride (SiN) nanopillars on a silica ($SiO_2$) substrate with different in-plane orientations $\theta$ (Fig. 3A). We numerically simulate the cross-polarization amplitude of the nanopillar of different lengths and widths (Fig. 3B). To ensure both high transmission efficiency and cross-polarization amplitude, we select one nanopillar with length and width of 285 nm and 150 nm, respectively. To verify the broadband performance of our geometric metasurface, we further simulated the cross-polarization amplitude of the nanopillar in the wavelength range from 400 nm to 700 nm (Fig. 3C). As a result, we designed and fabricated our wavelength-vortex multiplexing meta-hologram for the reconstruction of 36 independent image channels based on selected twelve wavelengths ($\lambda$ = 455 nm, 472 nm, 486 nm, 503 nm, 518 nm, 539 nm, 558 nm, 580 nm, 602 nm, 624 nm, 648 nm, 677 nm) and three OAM states ($l$ = -1, -3, -5)(*48*).

The experimental setup used for characterizing the performance of the wavelength-vortex multiplexing meta-hologram is illustrated in Fig. 3D(*48*). Vortex beams at different wavelengths were generated from a supercontinuum laser source after passing through vortex holograms imprinted on a spatial light modulator (SLM). Notably, to experimentally implement the KAA for *k*-space image filtering (fig. S4), we design a separate meta-hologram for the reconstruction of the 2D Dirac comb function following the same sampling constant as of the wavelength-vortex multiplexing hologram. As such, 36 independent images with Gaussian-like image pixels were experimentally reconstructed (Fig. 3E). To obtain the linewidth of the reconstructed light beams, the SSIMs of the demultiplexed images with respect to their target images were analyzed in a broadband (fig. S5). The average SSIM of the 36 multiplexed images is 0.82 for the incident light beams with correct wavelengths. The average linewidths for the 12 used wavelengths in the 36 image channels are 3.7 nm, 4.3 nm, 3.7 nm, 4.7 nm, 5 nm, 6.3 nm, 6.7 nm, 7 nm, 9.3 nm, 9.3 nm, 9.7 nm, and 10.7 nm, respectively (Fig. 3F).

**Wavelength-OAM-sharing Meta-holographic Visual Cryptography**

Visual cryptography (VC) inherits the idea of secret sharing and achieves secret reconstruction simply via the human visual system, which has recently been extended to the optical domain(*49*). In the VC scheme, a slight misalignment of digital share leads to failure of secret reconstruction, so ultranarrow-linewidth wavelength-vortex multiplexing meta-holography enables high-security colored metasurface HVC, in which both the OAM and wavelength information can be implemented as physical shares. During encryption, a colored secret image is divided into seven color blocks, with each block is further split into a pair of visual secrets(*48*) (Fig. 4A, fig. S6). Each visual secret is sampled by the 2D Dirac comb function with same sampling constant. Based on the encoded wavelengths ($\lambda$ = 436 nm, 466 nm, 497 nm, 531nm, 572 nm, 616 nm, 667 nm) and the OAM states ($l$ = 5, 3, 1, -1, -3, -5), a phase-only wavelength-vortex-multiplexing hologram used for the HVC application was generated using the transformer neural network. We then develop a metasurface hologram to encrypt the secret image, named as wavelength-OAM-sharing meta-ciphertext, with its physical shares (encoded in the OAM and wavelength degrees of freedom) being sent to 14 distinct shareholders.

During the decryption process, the holographic secret shares were reconstructed from the meta-ciphertext by using the correct wavelength-vortex beams (Fig. 4B). Notably, the spurious and undesired diffraction orders can be interpreted as noise preventing the disclosure of the overall secret image (fig. S7). Eventually, the shareholders can obtain the visual secrets with the KAA (Fig. 4B). The resultant final monochromatic or polychromatic secret images can be seen after the perfect spatial overlap of the separate holographic secret shares (Fig. 4C). The average linewidth of 5 nm suggests the precise requirement of the wavelength share information (Fig. 4D). To illustrate the anti-eavesdropping capabilities of our HVC, the experimental results of invalid decryption trials are given in figs. S8 and S9, showing that hackers with inappropriate OAM or wavelength physical shares fail to obtain the encrypted images. Further, the information rate, defined as the ratio of the secret size to the share size received by each shareholder, is analyzed to illustrate the security of HVC(*50*). In this context, the secret level is 7*49*49 bits, where 7 represents the wavelength channels and 49*49 represents the resolution of the binary gray-level secret. Each shareholder's received share level, represented by a binary sequence, is 13 bits (7 wavelength channels and 6 OAM channels). Specifically, in the binary sequence, 0 indicates that the current physical share is inaccessible, and 1 indicates it is accessible. Therefore, the information rate is 7*49*49/13 = 1293, which is more than 2500 times higher compared to the traditional VC scheme with an information rate of 1/2, demonstrating an unprecedented security.

**Conclusion and Outlook**

In summary, we have demonstrated ultranarrow-linewidth wavelength-vortex multiplexing holography in metasurface holograms. This is achieved by implementing elementary dispersion engineering of *k*-vector components. Using transformer neural networks for the design of phase-only multiplexing holograms, we showcased that 118 independent image channels can be reconstructed from a single metasurface hologram, achieving an ultranarrow linewidth of 2 nm in the visible range. Finally, we have employed the developed wavelength-vortex metasurface holograms for HVC, achieving unprecedented security with an information rate more than 2500 times higher than that of traditional visual cryptography schemes. We believe the upper limit of the multiplexing channel number is mainly constrained by the computational power of graphics processing unit (GPU, TITAN RTX 24GB in current scheme) for the hologram design. It is worthwhile mentioning that the principle of elementary dispersion engineering can be extended to 3D wavelength-vortex holography, wherein an additional Fresnel lens function can be added for considering image reconstruction in 3D (fig. S10). We believe our results open exciting avenues for the use of ultrathin holographic devices in various applications, including 3D displays(*51*), holographic encryption(*15*), beam shaping(*52*), LiDAR(*53*), microscopy(*54*), data storage(*3*), and optical artificial intelligence(*55*).

**References and Notes**

1. D. Gabor, A new microscopic principle. *Nature* **161**, 777–778 (1948).
2. D. Psaltis, D. Brady, X. G. Gu, S. Lin, Holography in artificial neural networks. *Nature* **343**, 325-330 (1990).

**Acknowledgments:**

**Funding:** X.F. acknowledges the funding support from the National Natural Science Foundation of China (62422509), the Shanghai Natural Science Foundation (23ZR1443700), the Shuguang Program of Shanghai Education Development Foundation and Shanghai Municipal Education Commission (23SG41), and the Young Elite Scientist Sponsorship Program by Cast (No.20220042). M. G. acknowledges the funding support from the Science and Technology Commission of Shanghai Municipality (Grant No. 21DZ1100500), the Shanghai Municipal Science and Technology Major Project, and the Shanghai Frontiers Science Center Program (2021–2025 No. 20). H.R. acknowledges funding support from the Australian Research Council (DE220101085, DP220102152). A.M. and J.F. acknowledge support from National Science Foundation (U.S.A.) grant 2127235. Part of this work was conducted at the Washington Nanofabrication Facility/Molecular Analysis Facility, a National Nanotechnology Coordinated Infrastructure (NNCI) site at the University of Washington, with partial support from the National Science Foundation via awards NNCI-1542101, NNCI-2025489. S.A.M. acknowledges funding support from the Australian Research Council (DP220102152) and the Lee Lucas Chair in Physics. This work was partially performed in part

## Supplementary Materials

Materials and Methods

Supplementary Text

Figs. S1 to S10

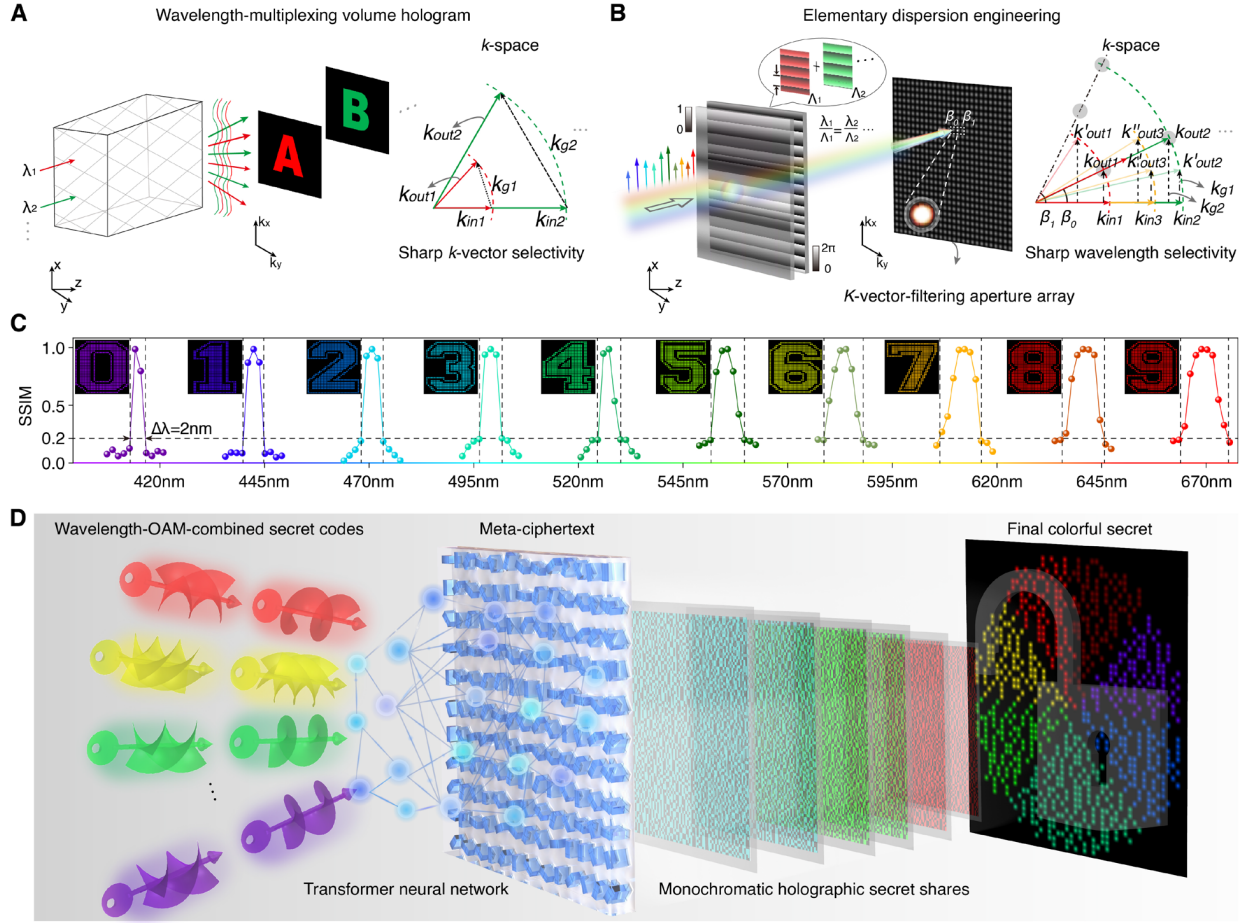

**Fig. 1. Conventional wavelength-multiplexing volume holography versus our strategy.** (**A**) Wavelength-multiplexing volume hologram. When the grating function vectors ($k_{g1}$, $k_{g2}$) are determined, the *k*-vectors of the incident light beams ($k_{in1}$ and $k_{in2}$) are highly selected due to Bragg diffraction (right panel). (**B**) Design of a complex-amplitude hologram comprising wavelength-dependent grating functions, associated with a sparse *k*-vector-filtering aperture array (KAA) in the momentum space. When the periods of the grating functions satisfy $\lambda_1/\Lambda_1 = \lambda_2/\Lambda_2$, the desired output light beams with the wavelengths of $\lambda_1$ and $\lambda_2$ can be guided into the aperture in KAA with a diffraction angle $\beta_0$ ($k_{out1}$ and $k_{out2}$). Meanwhile, KAA (e.g. with apertures in diffraction angles $\beta_0$, $\beta_1$) should block the other undesired wavelengths ($k'_{out3}$ and $k''_{out3}$) and spurious diffraction orders ($k'_{out1}$ and $k'_{out2}$). (**C**) Illustrational of ultranarrow-linewidth wavelength multiplexing holography using this scheme. The linewidth of the encoded reconstructive beams is represented as $\triangle\lambda$, which can reach 2 nm. (**D**) The wavelength-vortex multiplexing meta-holography is achieved when OAM is further introduced to increase the information channels. And a schematic of application in wavelength-OAM-sharing meta-holographic visual cryptography is given, wherein a transformer neural network is utilized to design the phase-only broadband meta-ciphertext for the wavelength-OAM-combined secret codes.

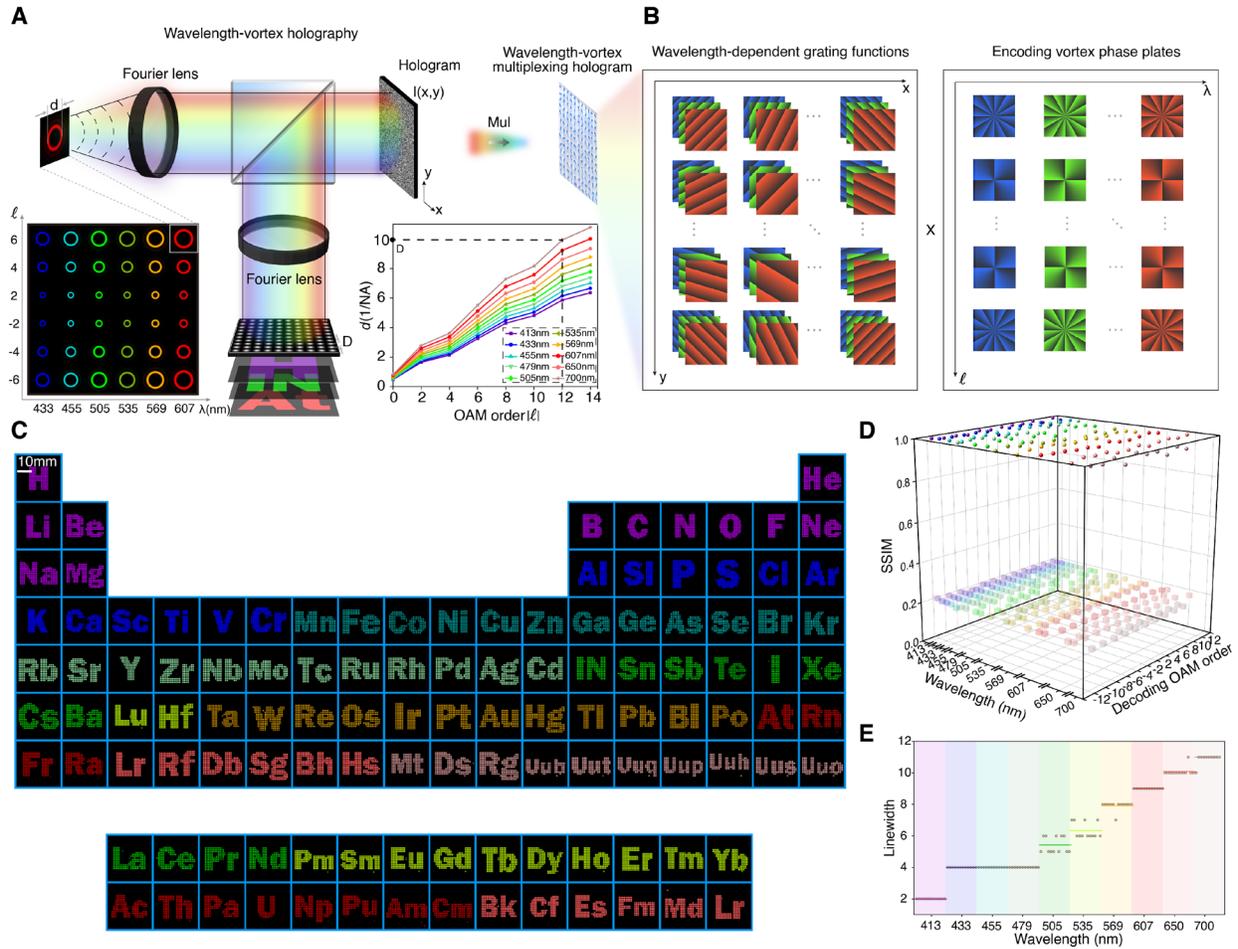

**Fig. 2. Design principle of ultranarrow-linewidth wavelength-vortex holography with high-capacity.** (**A**) Design principle of wavelength-vortex holograms. The spatial frequency of the vortex beams with different wavelengths is illustrated in bottom left, which determines the sampling constant of the 2D comb function shown in the bottom right. (**B**) The Fourier components of the wavelength-vortex multiplexing hologram comprising wavelength-dependent grating functions and encoding vortex phase plates. (**C**) 118 independent reconstructive images using the KAA. (**D**) SSIM analysis of the reconstructive images for the incident beams with distinct OAM states and wavelengths. (**E**) Linewidth analysis of the illuminated vortex beams with distinct multiplexed wavelengths.

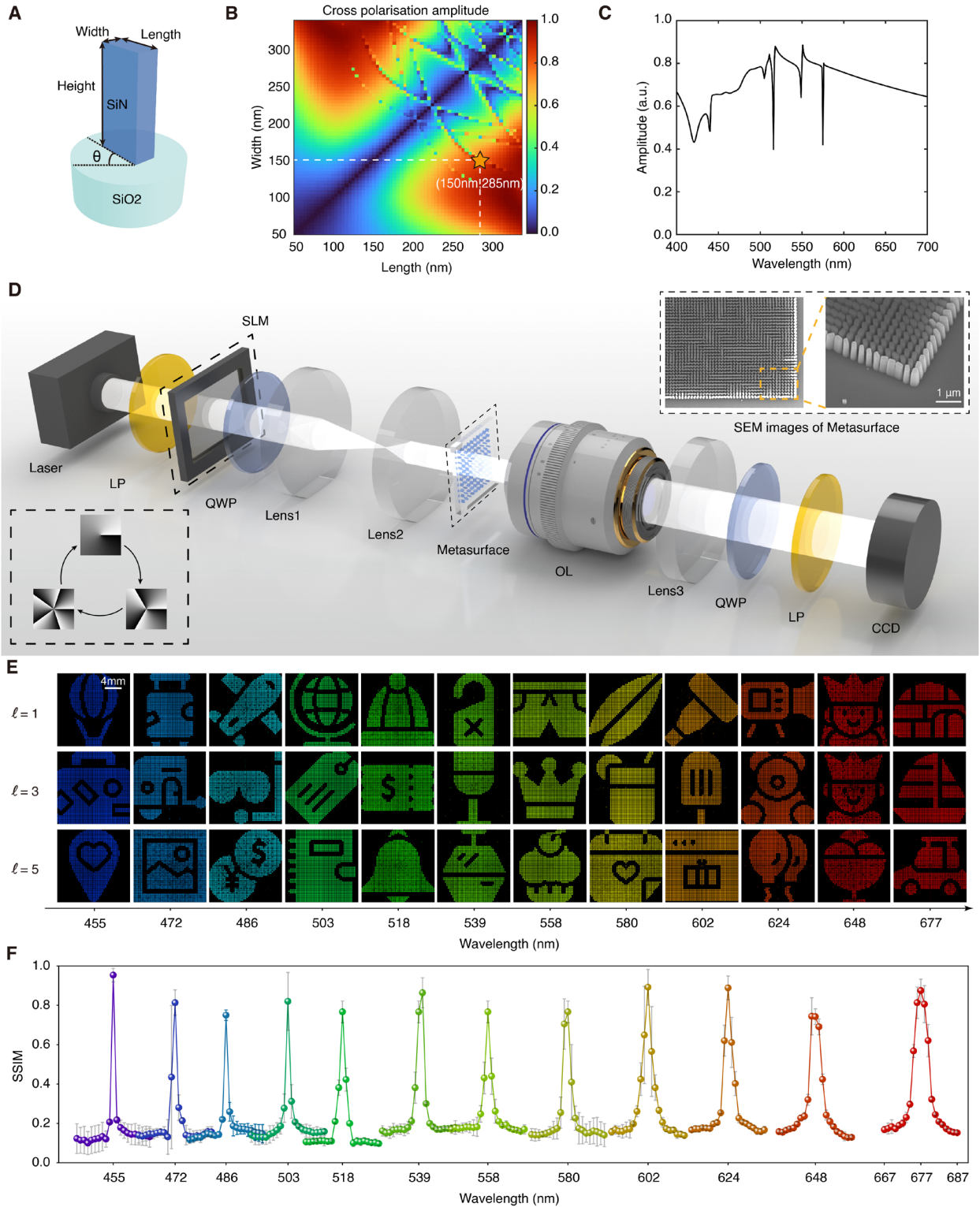

**Fig. 3. Experimental demonstration of the ultranarrow-linewidth wavelength-vortex multiplexing meta-holography.** (**A**) Schematic diagram of sub-wavelength SiN nanopillars on SiO$_2$ substrate. $\theta$ indicates the in-plane rotation angle of SiN nanopillars. (**B**) Numerical characterization of cross-polarization amplitude transmission after transmission through

nanopillars of different heights and widths. Selected nanopillar is marked with asterisks in the figure. (**C**) Numerical characterization of nanopillar amplitude transmission efficiency in the range of 400 to 700 nm. (**D**) Conceptual illustration of optical setup. LP: Linear polarizer, QWP: Quarter-wave plate, SLM: Spatial light modulator, OL: Objective lens, CCD: Charge-coupled device. (**E**) 36 independent reconstructive images. (**F**) SSIM analysis to obtain the linewidth of the multiplexed vortex beams.

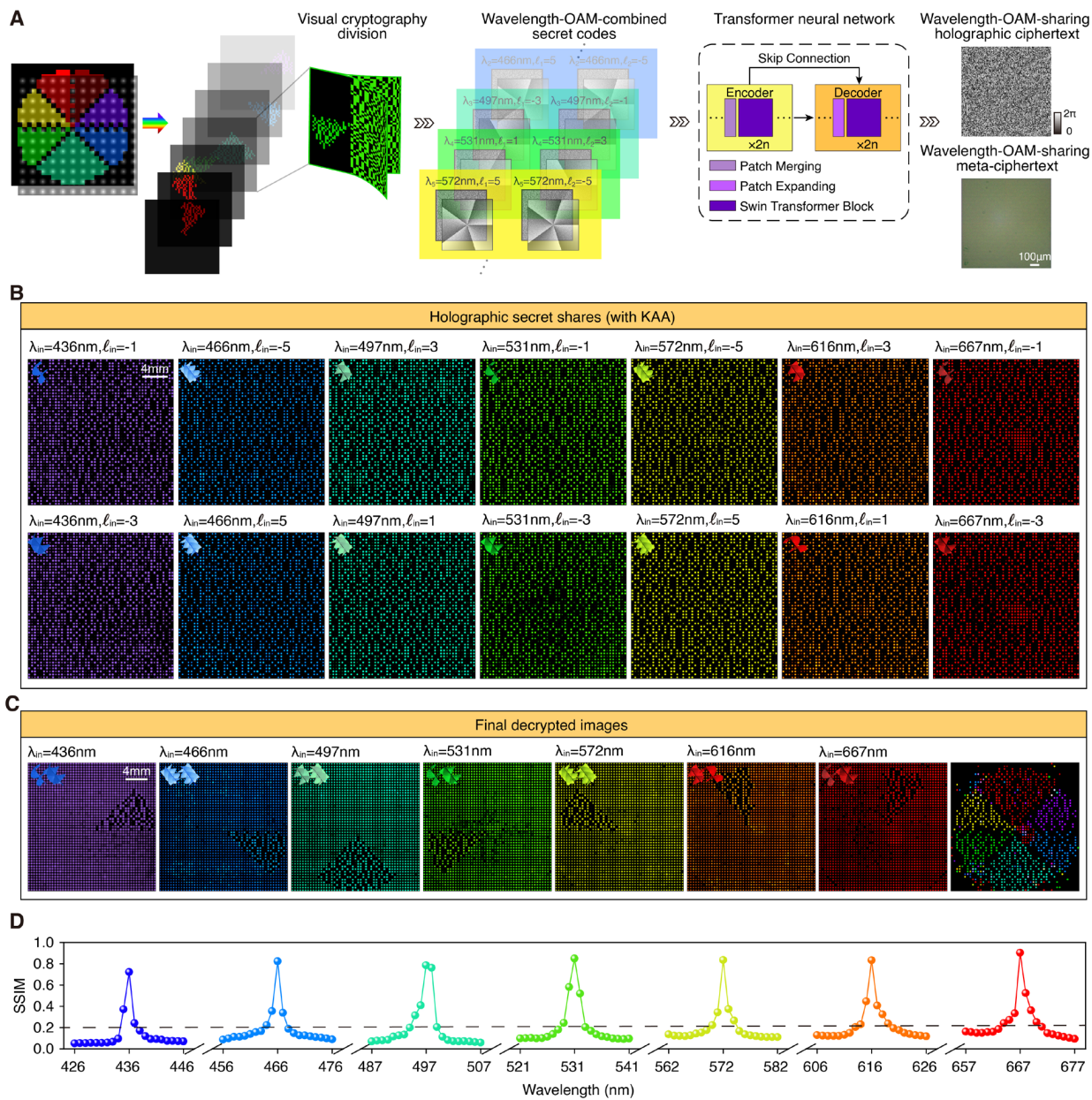

**Fig. 4. Experimental demonstration of Wavelength-OAM-sharing meta-holographic visual cryptography.** (**A**) Design principle of wavelength-OAM-sharing meta-ciphertext. (**B**) Reconstructive holographic secret shares with the KAA. (**C**) Reconstruction of the final monochromatic and overall chromatic secret images. (**D**) SSIM and linewidth analysis to illustrate the anti-eavesdropping of this strategy.

14